\documentclass[twocolumn,superscriptaddress]{revtex4}
\usepackage{graphicx}
\usepackage{subfig}
\usepackage{float}
\usepackage{bm}
\usepackage{color}
\usepackage{amsmath}
\usepackage{amsfonts}
\usepackage{amssymb}
\usepackage{natbib}
\usepackage{latexsym}
\usepackage{mathrsfs}

\def \ee{\end{equation}}
\def \be{\begin{equation}}
\def \eea{\end{eqnarray}}
\def \bea{\begin{eqnarray}}

\begin{document}

\title{Cosmological electromagnetic Hopfions}

\author{Sergio A. Hojman}
\email{sergio.hojman@uai.cl}
\affiliation{Departamento de Ciencias, Facultad de Artes Liberales,
Universidad Adolfo Ib\'a\~nez, Santiago 7491169, Chile.}
\affiliation{Departamento de F\'{\i}sica, Facultad de Ciencias, Universidad de Chile,
Santiago 7800003, Chile.}
\author{Felipe A. Asenjo}
\email{felipe.asenjo@uai.cl}
\affiliation{Facultad de Ingenier\'ia y Ciencias,
Universidad Adolfo Ib\'a\~nez, Santiago 7491169, Chile.}

\date{\today}

\begin{abstract}
 It is shown that  any  mathematical  solution for  null electromagnetic field knots in flat spacetime is also a  null field knotted solution for cosmological electromagnetic fields. This is  obtained by   replacing the  time $t\rightarrow \tau=\int dt/a$, where $a=a(t)$ is the scale factor of the Universe described by the Friedman-Lema\^itre-Robertson-Walker (FLRW)   cosmology, and by adequately rewriting the (empty flat spacetimes) electromagnetic fields solutions in a medium  defined by the FLRW metric. We found that the dispersion (evolution) of electromagnetic Hopfions is faster on cosmological scenarios. 
 We discuss the implications of these results for different cosmological models.
\end{abstract}



\maketitle

\section{Introduction}

In an article published in 1931, Hopf introduced a transformation \cite{hopf} that is nowadays referred to by physicists as {\it Hopf map} or {\it Hopf fibration} although, the same transformation was found by Hurwitz 33 years earlier in 1898 \cite{hurwitz}. This transformation is called {\it Hurwitz map} by  mathematicians. We will adhere to the physicist's way of naming it due to the fact that there are physical entities mathematically based on this transformation that are named {\it Hopfions}.
It is interesting to mention  that Dirac published in 1931 his very well known results on magnetic monopoles \cite{dirac}. Nowadays it is known that Dirac's magnetic monopoles may be described as Hopfions (see, for instance \cite{ryder}).

Although the main purpose of this work is to deal with electromagnetic Hopfions on both empty flat spacetime and on cosmological backgrounds, it is only fair to say that Hopf map can be found in physics in many different subjects ranging from fluid dynamics to superconductivity to quantum computing to electromagnetism to skyrmions. Some of these topics are discussed in the article {\it The Hopf fibration--seven times in physics} by Urbantke \cite{urbantke}. 
 In this way, hopfions can be found and studied in field theory \cite{foster}, condensed matter physics \cite{zzheng,Tikhonov,Ackerman2,szou}, magnetic systems \cite{rybakov,Sutcliffe,Reynolds,Liu}
gravitation \cite{shnir,amy}, optics \cite{cwan,yshen,Ehrmanntraut}, or
light \cite{particle,array2}, among several others.  
Furthermore, it is important to stress that physical Hopfions do exist, in the sense that they have been already produced in the laboratory in numerous experiments (see, for instance Refs.~\cite{Jung-Shen,particle,Ackerman,BGg,Reynolds,zzheng,Tikhonov,Ackerman2}).

The usual way to deal with Hopf map \cite{hopf} starts with the statement that it is defined as an injection from $\mathbb{R}^4$ with cartesian coordinates $u_1$, $u_2$, $u_3$, and $u_4$ to $\mathbb{R}^3$ with cartesian coordinates $x_1$, $x_2$, and $x_3$ defined by
\bea \label{hopf1}
x_1&=& 2(u_1 u_3+u_2 u_4)\ , \nonumber \\ 
x_2&=& 2(-u_1 u_4+u_2 u_3)\ ,\nonumber \\
x_3&=&{u_1}^2+ {u_2}^2-{u_3}^2- {u_4}^2\ ,
\eea
\noindent which preserves spheres, i.e., an  $\mathbb{S}^3$ sphere of radius $R$ in $\mathbb{R}^4$ is mapped on an $\mathbb{S}^2$ sphere of radius $r$ in $\mathbb{R}^3$ in such a way that
\be \label{rR}
r\equiv \sqrt{{x_1}^2+ {x_2}^2+{x_3}^2}={u_1}^2+ {u_2}^2+{u_3}^2+{u_4}^2 \equiv R^2\ ,
\ee
\noindent
as it  can be easily verified using Eqs.~\eqref{hopf1}.
It is not difficult to realize that Eqs.~\eqref{hopf1} may be inverted to get
\bea \label{hopf2}
u_1&=&\sqrt{r} \cos\frac{\theta}{2}\cos\frac{\psi+\phi}{2}\ , \nonumber \\ 
u_2&=&\sqrt{r} \cos\frac{\theta}{2}\sin\frac{\psi+\phi}{2}\ , \nonumber \\
u_3&=&\sqrt{r} \sin\frac{\theta}{2}\cos\frac{\psi-\phi}{2}\ ,\nonumber \\
u_4&=&\sqrt{r} \sin\frac{\theta}{2}\sin\frac{\psi-\phi}{2}\ ,
\eea
\noindent where $r$, $\theta$ and $\phi$ are the usual spherical coordinates related to cartesian ones by
\bea \label{csph}
x_1&=& r \sin \theta \cos \phi\ , \nonumber \\ 
x_2&=& r \sin \theta \sin \phi\ ,\nonumber \\
x_3&=&r \cos \theta \ ,
\eea
\noindent being $\psi$  arbitrary.

As a matter of fact, strictly speaking, Eq.\eqref{hopf1} does not have a proper (or unique) inverse in the sense that a map from $\mathbb{R}^3$ to $\mathbb{R}^4$ is not well defined, unless you introduce an arbitrary new variable ($\psi$) which plays the role of a gauge.

The gauge variable $\psi$ is sometimes referred to as the fiber of the Hopf map (in the fiber bundle sense). In this case, the fiber is a circle on the $\mathbb{S}^3$ sphere which is mapped on a single point on the $\mathbb{S}^2$ sphere. When one performs the inverse map, that single point on the $\mathbb{S}^2$ sphere back on the  $\mathbb{S}^3$ sphere, there is a multitude of pre--images defined by $0 \le \psi <2 \pi$, which is a circle.

We start this work with a short review of electrogmagnetic hopfions in Sec. II. Then, in Sec. III, we present the general formulation of Maxwell equations in Friedman-Lema\^itre-Robertson-Walker  (FLRW) cosmology, to later, in Sec. IV, present  a simple  Hopfion solution for general cosmology,  analyzing its behavior on different cosmological scenarios. Finally, in Sec. V we discuss these findings.

\section{Hopfions in Electromagnetism}

Hopfions are knotted topological solitons which are solutions to different kinds of field equations in physics. They are best described in terms of the Hopf map.

To the best of our knowledge, the first explicit electromagnetic Hopfions solutions to Maxwell equations in empty flat spacetime were published by A. F. Ra\~nada in 1989, as shown, for instance, in some of his articles \cite{rañada1,rañada2,rañada3,rañada4,rañada5}. It is fair to state that A. Trautman \cite{TRAUTMAN}  discovered the Hopfion existence in 1977 and presented it in a general abstract fashion without exhibiting explicit examples, that is probably the reason his article did not attract much attention until recently.

Ra\~nada started  by defining the initial ($t=0$) values of the electromagnetic fields $\vec{E} (x,y,z,0)$ and $\vec{B} (x,y,z,0)$ (which satisfy the initial value constraints) as solutions to the Hopf map. To get the time dependent fields $\vec{E} (x,y,z,t)$ and $\vec{B} (x,y,z,t)$ is sufficient to propagate the initial value fields using the Fourier transform to satisfy the full set of Maxwell equations.
The final form of Ra\~nada's electromagnetic Hopfion is given by 
\begin{eqnarray}\label{b}
\vec{B} (\vec{r},t)&=&\frac{\alpha b^2}{(2\pi)^{3/2}} \int  \vec{R} (\vec{k})\ \cos(\vec{k}\cdot\vec{r}-\omega t)\ d^3 \vec{k}  \nonumber \\
&+&\frac{\alpha b^2}{(2\pi)^{3/2}} \int  \vec{R}\ ' (\vec{k})\ \sin(\vec{k}\cdot\vec{r}-\omega t)\ d^3 \vec{k}\ ,
\end{eqnarray}
and
\begin{eqnarray}\label{e}
\vec{E} (\vec{r},t)&=&\frac{\alpha b^2}{(2\pi)^{3/2}} \int  \vec{R}\ ' (\vec{k})\ \cos(\vec{k}\cdot\vec{r}-\omega t)\ d^3 \vec{k}  \nonumber \\
&-&\frac{\alpha b^2}{(2\pi)^{3/2}} \int  \vec{R} (\vec{k})\ \sin(\vec{k}\cdot\vec{r}-\omega t)\ d^3 \vec{k}\ ,
\end{eqnarray}
where $\alpha$ and $b$ are constants with units of action and inverse of length, respectively. Also, $k$ and $\omega$ are the variables in the Fourier transform space. Besides,
\begin{equation}\label{r}
\vec{R} (\vec{k})=\frac{e^{- \omega}}{(2\pi)^{1/2} \omega} \Big(k_1 k_3,k_3(k_2+\omega),-k_1^2-k_2(k_2+\omega) \Big) \ ,
\end{equation}
and
\begin{equation}\label{r'}
\vec{R} \ '(\vec{k})=\frac{e^{- \omega}}{(2\pi)^{1/2} \omega} \Big(k_3^2+k_2(k_2+\omega),-k_1(k_2+\omega),-k_1k_3 \Big) \ .
\end{equation}

In the next section we will show that a slightly modified form of (empty flat spacetime) Ra\~nada's Hopfion given by Eqs.~\eqref{b}--\eqref{r'},  produces an electromagnetic cosmological Hopfion on a FLRW cosmological background. This cosmological electromagnetic Hopfions have a dynamical evolution that depends on the background cosmological Universe content. This work is in the same spirit of the results found for electromagnetic Hopfions in de Sitter spacetime \cite{Grzelaa}. However, in here, we present a study of a more general setting of different cosmological models, described by the FLRW metric.

The results may be obtained by taking advantage of the conformal flat nature of Maxwell equation under the FLRW metric. Thus, any Hopfion solution of flat spacetime Maxwell equation will be also a solution of Maxwell equations in cosmology, evolving on a cosmological time, and with analogue properties to the ones in flat spacetimes.

Below, we study a Hopfion in an expanding Universe filled by a perfect fluid with positive equation of state. Also, we explore the  form how a Hopfion disperses in a accelerated expanding Universe filled by a cosmological constant that plays the role of dark energy. In all these cases, we show that electromagnetic Hopfions evolve (and disperse) faster than in flat spacetime, as the cosmological time that determines the structure of the Hopfion runs faster than the time coordinate of flat spacetime.

\section{Maxwell equations in FLRW cosmology}

In curved spacetime, Maxwell equations are written as 
\begin{equation}
\nabla_\mu F^{\mu\nu}=0\, ,\quad \nabla_\mu F^{*\mu\nu}=0 \, ,
\label{Max1}
\end{equation}
in terms of covariant derivative operators for a  spacetime metric $g_{\mu\nu}$. Here,  $F_{\mu\nu}$ is the electromagnetic tensor  (with its dual $F_{\mu\nu}^*$).
We can define the vectorial electric fields as
$E_i=F_{i0}$, and $D^i=\sqrt{-g} F^{0i}$. Similarly, we define  the magnetic fields as $B^i=\varepsilon^{0ijk}F_{jk}$ and $\varepsilon^{0ijk}H_k=\sqrt{-g} F^{ij}$. They fulfill the relations \cite{Plebanksi,Felice,mass1,mass2,asenjohojman}
\bea
D^i&=&-\sqrt{-g}\left(\frac{g^{ij}}{g_{00}}\right)E_j+\varepsilon^{0ijk}\left(\frac{g_{0j}}{g_{00}}\right)H_k\, ,\nonumber\\
B^i&=&-\sqrt{-g}\left(\frac{g^{ij}}{g_{00}}\right)H_j-\varepsilon^{0ijk}\left(\frac{g_{0j}}{g_{00}}\right)E_k\, .
\eea

In the following we consider a cosmological model described by the FLRW metric 
\be
g_{\mu\nu}={\mbox{diag}}(-1,a^2,a^2,a^2)\, ,
\ee
(in Cartesian coordinates),
where $a = a(t)$ is the scale factor of the Universe, in terms of the proper time $t$.

In this case, we have
that 
\bea
{\bf D}&=&a {\bf E}\, ,\nonumber\\
{\bf B}&=&a {\bf H}\, ,
\eea
and cosmological curved spacetime Maxwell equations \eqref{Max1} can be simply written as
\begin{eqnarray}
\nabla\cdot {\bf B}&=&0\, ,\quad 
\nabla\cdot {\bf D}=0\, , \nonumber\\
 \frac{\partial\bf D}{\partial  \tau}&=&\nabla\times {\bf B}\, , \quad
\frac{\partial\bf B}{\partial \tau}=-\nabla\times {\bf D}\, , 
\label{Max2}
\end{eqnarray}
 in terms of the conformal time
\begin{equation}
    \tau=\int \frac{dt}{a}\, ,
\end{equation}
and flat spacetime gradient operator $\nabla$.

These equations are analogue to Maxwell equations in flat spacetime. Furthermore, we can define the Riemann–Silberstein vector for the FLRW system, $\textbf{\emph{F}}={\bf D}+i{\bf B}=a \left({\bf E}+i  {\bf H}\right)$, to re-write Eqs.~\eqref{Max2} as
\begin{eqnarray}
i\frac{\partial\textbf{\emph{F}}}{\partial  \tau}&=&\nabla\times \textbf{\emph{F}}\, .
\label{Max2b}
\end{eqnarray} 

The previous calculation shows that, as the FLRW cosmology is conformally flat, any solution of the source-free 
Maxwell equations
  in Minkowski spacetime will remain a solution in FLRW spacetime in the conformal time $\tau$. Furthermore, the energy $\int d^3r\, \textbf{\emph{F}}^*\cdot \textbf{\emph{F}}$ of any solution is conserved in the conformal time.

\section{Cosmological Hopfions}

In general, all flat spacetime Hopfions solutions of Maxwell equation will be solution of the cosmological FLRW Maxwell equations \eqref{Max2}, now for the conformal time.

Several null solutions of Maxwell equations are known for flat spacetimes (see, for example,  Refs.~\cite{rañada1,rañada2,rañada3,rañada4,rañada5,iwoff3, besieris, irvine,particle,array,TRAUTMAN,amy,irvine2,hoyos}). Due to the conformally flat nature of FLRW cosmology, every one of these solutions are also cosmological electromagnetic Hopfions solutions, with the corresponding change in time.
This is not a trivial change, as the conformal time $\tau$ depends on the cosmological Universe where electromagnetic fields develop. Thereby, electromagnetic Hopfions evolve different in different cosmological settings, compared to flat spacetime.

In here, we 
display this behavior by studying a simplest Hopfion solution \cite{besieris, irvine}.
The equivalent cosmological electrogmagnetic hopfion solution of Eqs.~\eqref{Max2} is
\begin{eqnarray}
{\bf D}=\frac{\nabla\zeta\times\nabla{\bar \zeta}}{4\pi i(1+{\bar\zeta}\, \zeta)^2}\, ,\quad
    {\bf B}=\frac{\nabla\eta\times\nabla{\bar \eta}}{4\pi i(1+{\bar\eta}\, \eta)^2}\, ,  
\label{hopfionsolu}
\end{eqnarray}
where (in Cartesian coordinates, and conformal time $\tau$)
\begin{eqnarray}
    \zeta(\tau,x,y,z)&=&\frac{A x+y\, \tau+i\left(A z+(A-1)\tau\right)}{x\, \tau-Ay+i\left(A(A-1)-z\, \tau \right)}\, ,\nonumber\\ \eta(\tau,x,y,z)&=&\frac{A z+(A-1)\tau+i\left(x\,  \tau-A y\right)}{A x+y\, \tau+i\left(A(A-1)-z\, \tau \right)}\, ,\nonumber\\
    A(\tau,x,y,z)&=&\left(x^2+y^2+z^2+1-\tau^2\right)/2\, ,    
\end{eqnarray}
and ${\bar\eta}$, ${\bar \zeta}$ are the  complex conjugated of such functions. 
This hopfion solution fulfills
\bea
{\bf B}\cdot{\bf D}&=&0={\bf H}\cdot{\bf E}\, ,\nonumber\\
|{\bf B}|^2&=&|{\bf D}|^2=a^2|{\bf H}|^2=a^2|{\bf E}|^2\, .
\eea

As we will see, the nonlinearity of the $t \rightarrow\  \tau$ transformation 
produces significant changes in the evolution of Hopfions.
A Universe filled with perfect fluid, with a proper energy to pressure  ratio given by $w$, has a scale factor  given by  \cite{ryden}
\be
a(t)=\left(\frac{t}{t_0}\right)^{2/(3+3w)}\, ,
\ee  
with $0<t<t_0$, where $t_0$   is the current age of the Universe. Thus, for this case, the conformal time becomes
\begin{equation}
\tau=\frac{3(1+w)\, t}{1+3w}\left(\frac{t_0}{t} \right)^{{2}/({3+3w})}\, .
    \label{tautau}
\end{equation}
In here, we consider only normal fluids
with $0<w<1$. 
Due to  this, then
$\tau>t$ always. Therefore, for these cases,  the Hopfions evolve faster
in these cosmologies.

Furthermore, for dark energy cosmology owe to a cosmological constant $\Lambda$, the scale factor becomes 
\be 
a(t)=\exp[H_0(t-t_0)]\, ,
\ee
where $H_0=\sqrt{\Lambda/3}$. The conformal time for this model is
\begin{eqnarray}
    \tau=\frac{1}{H_0}e^{H_0 t_0}\left(1- e^{-H_0 t}\right)\, ,
\end{eqnarray}
which again runs faster than $t$.

In order to depict  the evolution of Hopfions in these different scenarios we plot solution  
\eqref{hopfionsolu} for three different times in flat spacetime, and for the corresponding equivalent times for a radiation-dominated Universe (with $w=1/3$, and $\tau=2\sqrt{t_0 t}$), a matter-dominated Universe (with $w=0$, and $\tau=(3 t_0^2\, t)^{1/3}$), and a dark energy cosmology.
 
\begin{figure*}[ht]
\begin{minipage}{.5\linewidth}
\centering

\subfloat[$t=0$]{\includegraphics[width = 2.8in]{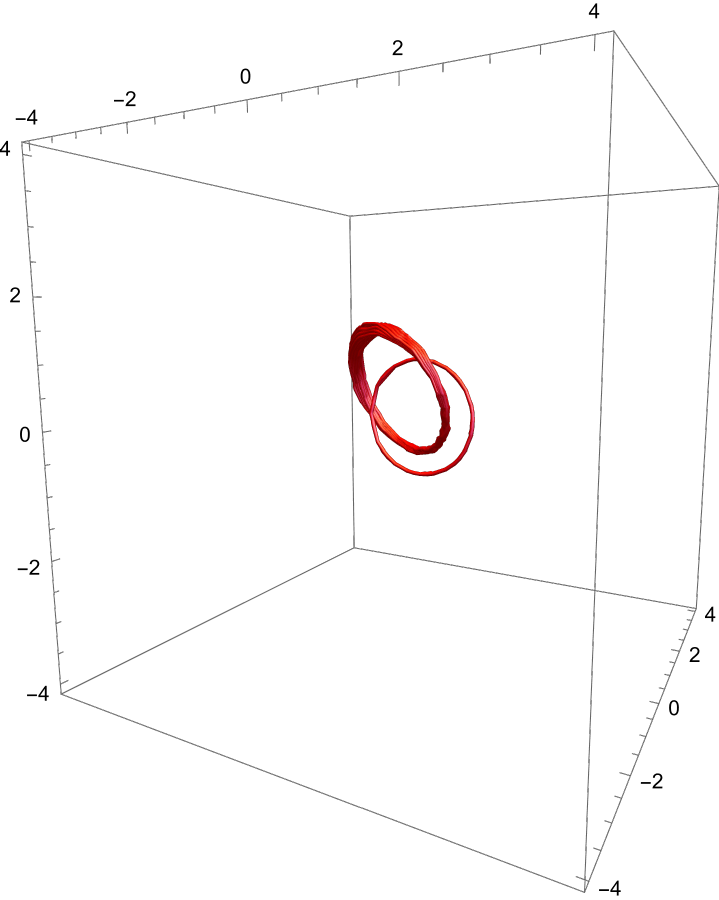}} 
\end{minipage}%
\begin{minipage}{.5\linewidth}
\centering
\subfloat[$t/t_0=1/2$] {\includegraphics[width = 2.8in]{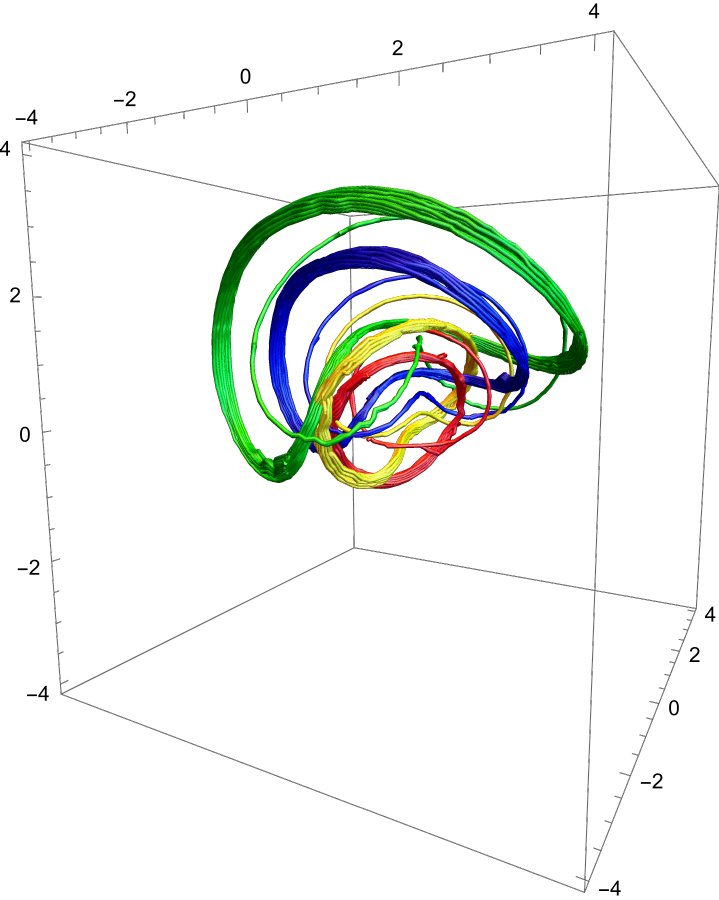}}
\end{minipage}\par\medskip
\centering
\subfloat[$t/t_0=1$] {\includegraphics[width = 2.8in]{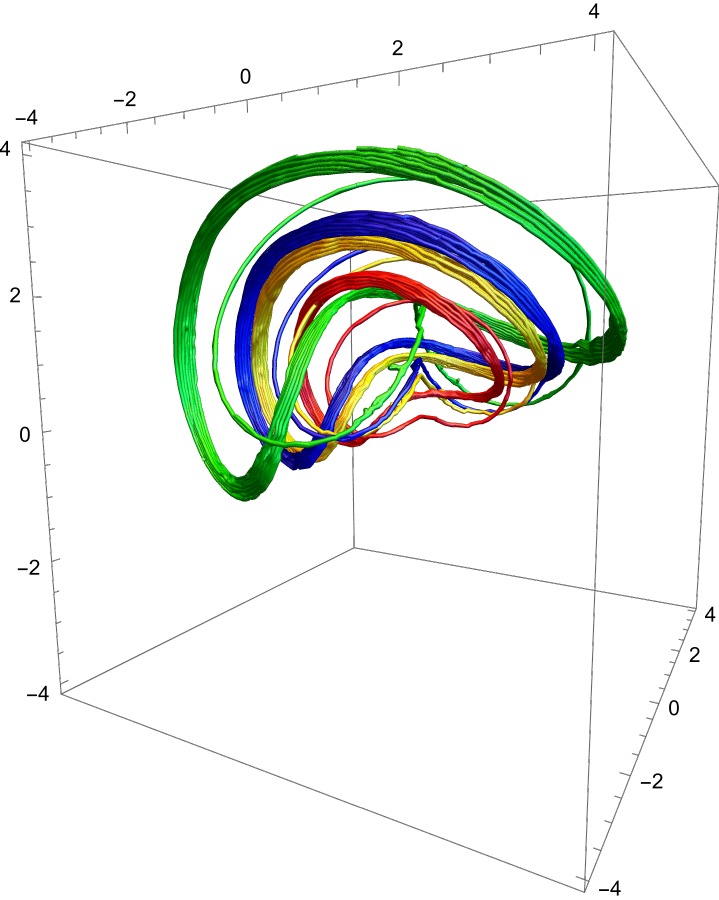}}
\caption{Evolution of hopfion \eqref{hopfionsolu} of field lines ${\bf D}$  in different cosmological settings. (a) For $t=0$. (b) For $t/t_0=1/2$, flat spacetime hopfion (red), radiation-dominated hopfion (blue),  matter-dominated hopfion (green), and  dark-energy cosmological hopfion (yellow). (c) For $t/t_0=1$ (same colors).}
\label{figura1}
\end{figure*}

In Fig.~\ref{figura1}, we plot the field lines of
the hopfion solution \eqref{hopfionsolu} for ${\bf D}$
for several cosmological models, in different coordinate times $t$ (all plot boxes have the same scales and orientations).
The hopfion solution for $t=0$ (with $a=1$), Fig. 1(a), coincide for all cosmological models.

In Figs. 1(b), we plot the different evolution for different cosmological models for $t/t_0=1/2$. In red lines, we plot the flat spacetime solution for comparison. These coincide with known solutions \cite{besieris, irvine}. In blue lines is depicted the same hopfion solution for a radiation-dominated scenario, corresponding to $\tau/t_0=\sqrt{2}$. In green lines, the hopfion solution for a matter-dominated cosmological model, for $\tau/t_0=(27/2)^{1/3}$.
Lastly, in yellow lines, we present the hopfion for dark energy cosmology, with $\tau/t_0=e-e^{1/2}$, where we have considered $H_0 t_0=1$.

In Figs. 1(c), we plot the solution for coordinate time $t/t_0=1$. Red lines are  the hopfion solution in flat spacetime. 
The hopfion solution in the radiation-dominated Universe, in blue lines, now is plotted for $\tau/t_0=2$. In green lines, the matter-dominated hopfion solution is obtained for $\tau/t_0=3$. Finally, the evolution in a dark-energy cosmology, in yellow lines, is  shown for $\tau/t_0=e-1$, with $H_0 t_0=1$.

Thereby, in Fig. 1, we can see how, for the same coordinate times, the hopfion evolves faster in a radiation-dominated Universe than in flat spacetime. Even faster evolution occurs for a matter-dominated Universe. Also, in a dark-energy scenario, faster development occur for the hopfion solution, which depends on the value of $H_0$.

These different evolution (spreading) rates of the cosmological Hopfions are related to the fact that any electromagnetic wavepacket spread with time. In the cosmological context, the spreading of the electromagnetic wavepacket can be calculated as \cite{iwoff,iwoff2}
\begin{equation}
    \frac{d^2}{d \tau^2}\int d^3r\, \left( r- \langle r(\tau)\rangle \right)^2 \textbf{\emph{F}}^*\cdot \textbf{\emph{F}}= 2\int d^3r\,  \textbf{\emph{F}}^*\cdot \textbf{\emph{F}}\, ,
    \label{spreddd}
\end{equation}
which can be proved directly from Eqs.~\eqref{Max2} for the center of the momentum frame of the electromagnetic wavepacket \cite{iwoff,lekner}. As the right-hand side of Eq.~\eqref{spreddd} is the conserved energy, this implies that the second conformal-time derivative of the spreading of an electromagnetic wavepacket is constant.
Thus, the spreading of a cosmological electromagnetic hopfion is proportional to the square of conformal time, $\tau^2$. In terms of the  coordinate time, this implies that Hopfions spread at different rates depending on the cosmological setting in which  they  develops.

\begin{figure}
{\includegraphics[width = 3.4in]{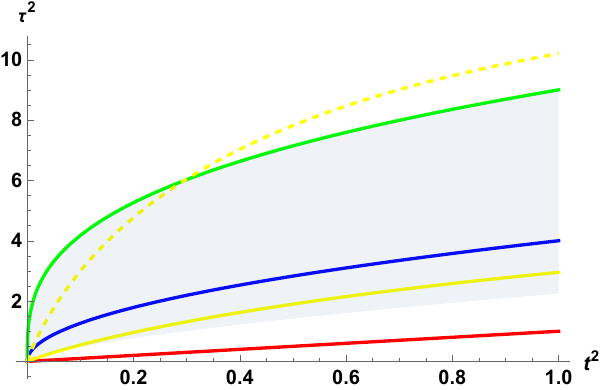}}
\caption{Cosmological time dependence of the electromagnetic wavepacket spreding with respect to its flat spacetime counterpart. Red line is for flat spacetime scenario. Blue and green lines are the faster spreading induced by the radiation-dominated and matter-dominated cosmologies, respectively. Yellow lines are for a Universe with dark energy content.}
\label{figura2}
\end{figure}

This dependence is shown in Fig.~\ref{figura2}, where we have plotted the square of the conformal time $\tau^2$ as a function of square of the proper time $t^2$. As expected, in flat spacetime $\tau^2=t^2$, which is shown as a straight red line. However, in any other cosmological scenario, the spreading is faster in terms of proper time. Besides, we show a shaded region that
 represents the different $\tau^2$ dependence in a perfect fluid cosmological setting, given by Eq.~\eqref{tautau},  for $0\leq w \leq 1$. 
 For all these cases, the spreading is faster. In blue line, we present the spreading rate for a radiation-dominated scenario, and similarly, for a matter-dominated cosmology in green line. Lastly, in yellow line, we show the spreading time $\tau^2$ for a dark-energy cosmology for $H_0 t_0=1$, and in yellow dashed-line for $H_0 t_0=2$. 

\section{Remarks}

All  flat spacetime electromagnetic Hopfions are also null electromagnetic solutions in  FLRW cosmology in a conformal time. Although this is a simple  consequence of the  conformally flat nature of the FLRW metric, the evolution of the hopfion solution  is different from the flat spacetime one, as the conformal time depends in a nonlinear way on the   coordinate time.

In the above studied cosmological scenarios, Hopfions  disperse faster than their counterparts in flat spacetime.  This can be seen in Fig. 1, where the Hopfions disperse as they evolve and depart from its initial doughnut--like shape. However, a  major general consequence of the cosmological evolution of any hopfion is the $\tau^2$-form of dispersion. This is valid for all cosmological electromagnetic solutions. Thus,  any other solution for cosmological Hopfions will have a faster dispersion $\tau^2>t^2$, compared to flat spacetime ones, as long as $\tau>t$. 

Nevertheless, this behavior is not valid for exotic forms of 
matter. For fluid with negative pressure (forms of dark energy), it is possible to have slower dispersion, such that $\tau^2<t^2$. Indeed, from Eq.~\eqref{tautau} we find that this occur for equations of state $-1<w\leq -2/3$. These $w$ define
a broad class of dark energy cosmologies without cosmological constant. The study of this class of cosmological electromagnetic Hopfions is left for the future.

The presented solution shows some features that can be observed in any cosmological hopfion.
But this solution also show a guideline on how more general hopfion-like solutions can be  possible to be obtained in cosmological settings for abelian fields, such as  heliknotons, skyrmions or torons \cite{WuSma}. 
Also, the presented cosmological electromagnetic hopfions experience similar effects than their flat spacetime counterparts, such as the possibility of  
change the topology, the energy distribution or the angular momenta when two  hopfions collide \cite{array2}, implying that cosmological hopfions can be important 
for the dynamics of the background radiation of  primordial Universe. On the other hand, the conformal nature of the presented hopfions allow to measure and model their characteristics in laboratory using their flat spacetime analogues \cite{Jung-Shen,particle}. Finally, the connection of cosmological electromagnetic hopfions (and of the other more general hopfion-like structures)  with quantum cosmological models with non-trivial geometry will be explored in the future \cite{ZeldovichZeldovich, Abarghouei}.

\begin{acknowledgements}
The authors thanks to I. Bialynicki-Birula for his guidance on some parts of this work.
FAA thanks to FONDECYT grant No. 1230094 that partially supported this work. 
 \end{acknowledgements}

\end{document}